\documentclass[journal]{IEEEtran}
\newcommand{\RNum}[1]{\uppercase\expandafter{\romannumeral #1\relax}}
\usepackage{mathrsfs}
\usepackage{amssymb}
\usepackage{diagbox}
\usepackage{cite}
\usepackage[T1]{fontenc}
\usepackage{graphicx}
\usepackage{algorithmic}
\usepackage[caption=false,font=footnotesize]{subfig}
\usepackage{makecell}
\usepackage{url}
\usepackage{stfloats}
\usepackage{hyperref}
\usepackage{amsmath}
\usepackage{array}
\usepackage[ruled,vlined,linesnumbered]{algorithm2e}
\usepackage{float}
\usepackage{multirow}
\usepackage{amsthm}
\usepackage{color}
\usepackage{multicol}
\usepackage{enumerate}
\usepackage{booktabs}  

\allowdisplaybreaks[4]

\ifCLASSINFOpdf

\else

\fi

\begin{document}

\title{Plasma GraphRAG: Physics-Grounded Parameter Selection for Gyrokinetic Simulations}

\author{Ruichen Zhang, Feda AlMuhisen, Chenguang Wan, Zhisong Qu, Kunpeng Li, Youngwoo Cho, Kyungtak Lim, Virginie Grandgirard, and Xavier Garbet

\thanks{
This research is supported by the National Research Foundation, Singapore. This research is also supported by Seatrium New Energy Laboratory, Singapore Ministry of Education (MOE) Tier 1 (RT5/23 and RG24/24), the Nanyang Technological University (NTU) Centre for Computational Technologies in Finance (NTU-CCTF), and the Research Innovation and Enterprise (RIE) 2025 Industry Alignment Fund - Industry Collaboration Projects (IAF-ICP) (Award I2301E0026), administered by Agency for Science, Technology and Research (A*STAR). (Corresponding author: Xaiver Garbet)}

\thanks{Ruichen Zhang and Kunpeng Li are with the School of Physical and Mathematical Sciences, Nanyang Technological University, Singapore, and also with the College of Computing and Data Science, Nanyang Technological University, Singapore (e-mail: ruichen.zhang@ntu.edu.sg; kunpeng.li@ntu.edu.sg). }

\thanks{Feda AlMuhisen and Virginie Grandgirard are with CEA, IRFM, F-13108 Saint Paul-lez-Durance, France (e-mail: virginie.grandgirard@cea.fr;  feda.almuhisen@cea.fr).}

\thanks{Chenguang Wan, Zhisong Qu, Youngwoo Cho, and Kyungtak Lim are with the School of Physical and Mathematical Sciences, Nanyang Technological University, Singapore. (e-mail: chenguang.wan@ntu.edu.sg; zhisong.qu@ntu.edu.sg; youngwoo.cho@ntu.edu.sg; kyungtak.lim@ntu.edu.sg).}

\thanks{
Xaiver Garbet is with the School of Physical and Mathematical Sciences, Nanyang Technological University, Singapore, and also with CEA, IRFM, F-13108 Saint Paul-lez-Durance, France (e-mail: xavier.garbet@ntu.edu.sg). }
}
\maketitle

\begin{abstract}
Accurate parameter selection is fundamental to gyrokinetic plasma simulations, yet current practices rely heavily on manual literature reviews, leading to inefficiencies and inconsistencies. We introduce Plasma GraphRAG, a novel framework that integrates Graph Retrieval-Augmented Generation (GraphRAG) with large language models (LLMs) for automated, physics-grounded parameter range identification. By constructing a domain-specific knowledge graph from curated plasma literature and enabling structured retrieval over graph-anchored entities and relations, Plasma GraphRAG enables LLMs to generate accurate, context-aware recommendations. Extensive evaluations across five metrics, comprehensiveness, diversity, grounding, hallucination, and empowerment, demonstrate that Plasma GraphRAG outperforms vanilla RAG by over $10\%$ in overall quality and reduces hallucination rates by up to $25\%$. {Beyond enhancing simulation reliability, Plasma GraphRAG offers a methodology for accelerating scientific discovery across complex, data-rich domains.}
\end{abstract}


\section{Introduction}

Understanding turbulence and transport in magnetically confined plasmas remains one of the central challenges in fusion energy research \cite{candy2004local}. Gyrokinetic (GK) simulations are indispensable for modeling these multiscale phenomena, as they capture the kinetic behavior of charged particles while reducing the dimensionality of the full six-dimensional Vlasov--Maxwell system \cite{lee1987gyrokinetic}. {Modern GK codes are widely used to investigate plasma stability, turbulence-driven transport, and performance limits in devices ranging from tokamaks to stellarators \cite{garbet2010gyrokinetic}.} These tools form the computational backbone of fusion research, providing insight into confinement optimization, transport barrier formation, and scenario development for future reactors.

A critical prerequisite for accurate GK simulations is the {selection of appropriate input parameter ranges}, including normalized temperature and density gradients, safety factors, magnetic shear, and collisionality. These parameters have a decisive impact on turbulence characteristics and transport predictions, and their accurate specification is essential not only for predictive modeling but also for the construction of surrogate models and databases. Traditionally, identifying suitable parameter ranges has relied on expert judgment and manual reviews of experimental and theoretical literature. This manual process is time-consuming, error-prone, and difficult to reproduce, {which results in} inconsistencies across benchmarking studies and undermine confidence in simulation outcomes.

Recent progress in {artificial intelligence (AI)}, particularly large language models (LLMs), has opened new possibilities for automating knowledge extraction from unstructured scientific literature \cite{10172151}. An LLM is a type of deep learning model trained on massive corpora to capture statistical patterns of language, enabling it to perform tasks such as summarization, reasoning, and question answering. To enhance domain specificity, LLMs are often combined with Retrieval-Augmented Generation (RAG), a paradigm where relevant documents are first retrieved from a knowledge base and then supplied as context to the LLM during generation \cite{gao2023retrieval}. While RAG has demonstrated effectiveness in many fields, standard implementations treat the literature as a flat corpus, failing to capture the \textit{structured interdependencies among physical variables} that are central to plasma physics \cite{10531073}. This limitation often leads to hallucinations or incomplete recommendations, particularly in highly technical applications such as gyrokinetic modeling.

To address these challenges, we introduce \textbf{Plasma GraphRAG}, a novel framework that integrates Graph Retrieval-Augmented Generation (GraphRAG) with LLMs to automate the identification of physics-grounded parameter ranges for gyrokinetic simulations. Plasma GraphRAG builds a domain-specific knowledge graph, where nodes represent plasma parameters, device metadata, and bibliographic sources, while edges encode physical couplings and co-occurrence relations. Structured retrieval over this graph provides the LLM with explicit relational context, enabling accurate, transparent, and reproducible parameter recommendations while significantly mitigating hallucinations\footnote{In this study, hallucination refers to cases where the model generates information that is inconsistent with the retrieved evidence or established physical principles.
A lower hallucination rate therefore indicates higher factual consistency and physical reliability in the model’s responses.}. The key contributions of this study are summarized as follows:
\begin{itemize}
    \item  We introduce Plasma GraphRAG, which constructs a domain-specific graph for GK modeling by harmonizing diverse literature sources into a unified, code-facing parameter space. By adopting standardized gyrokinetic notation and explicitly linking parameters with bibliographic evidence, the graph resolves long-standing inconsistencies across simulation codes and provides a reproducible foundation for parameter integration.
    
    \item  Plasma GraphRAG employs a retrieval mechanism that encodes physical couplings and co-occurrence relations among plasma parameters. Compared to standard RAG, this structured retrieval provides the LLM with richer context, enabling more accurate parameter extraction and improved interpretability through transparent evidence paths, while substantially reducing hallucinations.
    
    \item  We evaluate Plasma GraphRAG on controlled GK parameter identification benchmarks, showing improved response quality, diversity, and grounding relative to baseline methods. The framework streamlines the preparation of surrogate model databases, alleviates expert workload, and offers a scalable pathway toward reproducible, data-driven discovery in plasma turbulence and transport studies. {However, the current benchmark remains limited in scope, and the evaluation metrics are primarily heuristic, suggesting that larger-scale and quantitatively validated studies will be essential in future work.}
\end{itemize}

\section{Related Work}

This section reviews prior work in two relevant areas: (i) GK plasma physics and turbulence modeling, and (ii) LLMs with retrieval-based reasoning. The former outlines the evolution of GK simulation tools and surrogate modeling techniques, while the latter focuses on RAG and GraphRAG approaches for scientific question answering.

\subsection{Gyrokinetic Plasma Physics}

GK theory has been the cornerstone of turbulence and transport modeling in magnetically confined plasmas for several decades. Foundational work has established the nonlinear GK formalism for low-frequency waves, which underpins modern turbulence simulations \cite{frieman1982nonlinear}. {Since then, dedicated \emph{local} solvers such as \texttt{GS2}, \texttt{GENE}, \texttt{CGYRO}, and \texttt{GKW} have been developed and widely adopted to model ion- and electron-scale turbulence, trapped electron modes, and zonal flow dynamics \cite{jenko2000electron, candy2016high}. In parallel, \emph{global} gyrokinetic codes such as \texttt{GYSELA}, \texttt{ORB5}, and \texttt{GT5D} have been introduced to simulate turbulence and transport across the entire tokamak volume, capturing finite-radius effects and global profile variations \cite{donnel2019multi}.} These codes have enabled benchmarked studies of core transport phenomena, including the Dimits shift, and are now integral to scenario development for fusion devices. In parallel, reduced-order models such as Trapped Gyro-Landau Fluid (TGLF) \cite{staebler2007theory} and quasilinear models like QuaLiKiz \cite{garbet2010gyrokinetic} have been introduced to bridge detailed GK physics and integrated transport modeling. These models have achieved substantial speedups at the cost of some fidelity, enabling broader exploration of design spaces.

More recently, the field has embraced hybrid and data-driven modeling to address computational bottlenecks. Neural network surrogates trained on large GK datasets \cite{candy2016high} have been developed to provide flux predictions several orders of magnitude faster than first-principles simulations. Generative models \cite{clavier2025generative} and transformer-based 5D surrogates \cite{galletti20255d} have further enabled direct emulation of nonlinear turbulence with preserved spatiotemporal dynamics. Multi-fidelity methods have been proposed to combine reduced models with sparse high-fidelity GK data for improved predictive accuracy \cite{maeyama2024multi}. Ongoing benchmarking efforts have continued to validate solvers such as \texttt{GENE} and \texttt{CGYRO} under emerging physics regimes \cite{kim2024verification}. This evolution reflects a growing trend toward surrogate-augmented, data-driven workflows for turbulence-informed fusion scenario design.

\subsection{LLMs with Retrieval}

LLMs have demonstrated impressive capabilities across a range of NLP tasks, but they continue to struggle with factual accuracy and domain specificity. To address these limitations, retrieval-augmented generation (RAG) has emerged as a promising approach. Early systems such as DrQA \cite{chen2017reading} and REALM \cite{guu2020retrieval} have shown that integrating external retrieval with neural models improves factual grounding in open-domain question answering. The formal RAG framework introduced by \cite{lewis2020retrieval} has combined dense vector retrievers with encoder-decoder models like BART and T5 to produce contextualized answers using retrieved documents. However, standard RAG approaches have treated the underlying corpus as a flat collection of documents, ignoring structured relationships such as scientific couplings, units, or hierarchical parameter dependencies. This design choice has limited their effectiveness in technical domains, where relational reasoning and symbolic consistency are critical.

To address these shortcomings, recent work has explored GraphRAG, a class of models that incorporate structured knowledge representations {\cite{10771030}}. These methods have embedded queries into graph spaces, retrieved subgraphs composed of entities and their relations, and linearized the result into evidence paths for LLMs \cite{peng2024graph}. Applications have spanned enterprise QA, manufacturing documents, and e-commerce workflows \cite{knollmeyer2025document}, demonstrating improved interpretability and accuracy. In the scientific domain, systems such as PaperQA \cite{lala2023paperqa} have adapted RAG to scholarly literature, yielding enhanced factual coverage and traceable citations. Benchmarking studies \cite{han2025rag} have reported that GraphRAG excels at multi-hop reasoning and relationship synthesis. Nonetheless, challenges remain. Graph construction can introduce noise, especially in heterogeneous corpora, and task-specific graph schemas must often be hand-designed. Survey work \cite{ji2023survey} has highlighted hybrid architectures that combine both flat and graph-based retrieval to balance coverage, precision, and system complexity. Taken together, the above work suggests that GraphRAG can provide a promising foundation for LLM-based reasoning in structured scientific domains such as plasma physics. In this context, we propose Plasma GraphRAG, a domain-specific GraphRAG framework tailored to GK simulations. Our method leverages structured graphs to support reproducible, well-grounded parameter recommendations derived from literature evidence, addressing both domain complexity and LLM hallucination risk.

\section{Plasma GraphRAG Framework}
This section presents the architecture of Plasma GraphRAG, our proposed framework for automated, physics-grounded parameter range identification in GK simulations.

\subsection{Data Collection and Physics-Grounded Preprocessing}
The first stage of Plasma GraphRAG is the construction of a physics-grounded corpus tailored to GK parameter identification. This step defines the scope and quality of downstream knowledge graph construction and retrieval. We curate a structured dataset focused on descriptors that are standard in core transport and GK code-comparison studies, following the normalized notation of Bourdelle \emph{et al.}~\cite{bourdelle2015core}. At this stage, no numerical thresholds or scan protocols are fixed, and those are instantiated later during benchmark evaluation.

{The variable families span \textbf{geometry and magnetic equilibrium} 
(\( q,\, s,\, R_0/a,\, r/a \)), 
\textbf{thermodynamics and composition} 
(\( T_i/T_e,\, Z_{\mathrm{eff}} \)), 
\textbf{transport-driving gradients} 
(\( R/L_n,\, R/L_{T_e},\, R/L_{T_i} \)), 
and \textbf{kinetic or stability proxies} 
(\( \gamma_{E{\times}B}(a/c_s),\, \nu_{ei}(a/c_s) \)) 
used in core turbulence analyses~\cite{rodriguez2022nonlinear}. 
These variables are mapped into a standardized feature space, as defined in Eq.~(\ref{eq:gkfeatures}), 
ensuring a consistent input--output interface with GK codes and modeling pipelines.} \cite{rodriguez2022nonlinear}. These variables are mapped into a standardized feature space ${\mathcal{X}_{\mathrm{GK}}}$, i.e.,
\begin{equation}
\mathcal{X}_{\mathrm{GK}} =
\left\{
\begin{aligned}
 & q,\, s,\, R_0/a,\, r/a,\\
 & T_i/T_e,\, Z_{\mathrm{eff}}, \\
 & R/L_n,\, R/L_{T_e},\, R/L_{T_i},\\
 & \gamma_{E{\times}B}(a/c_s),\, \nu_{ei}(a/c_s),\\
 & [\,\beta_e,\, \rho_\ast\,]\,\,\,\,\,\,\,(\rm{Optional})
\end{aligned}
\right\},
\label{eq:gkfeatures}
\end{equation}
where additional dimensionless quantities such as $\beta_e$ and $\rho_\ast$ are included when available. This harmonized feature set ensures consistent input/output interface with GK codes and modeling pipelines.

To ensure physical integrity of the dataset, a harmonization process is applied that isolates quasi-steady-state core intervals, reconciles units and normalization schemes across sources, and removes tuples that are either incomplete or inconsistent with standard GK usage. This process yields a clean subset, i.e.,
\begin{equation}
D_{\mathrm{clean}}
= \bigl\{\, x \in D \;\big|\;
\mathcal{V}_{\mathrm{GK}}\!\left(x;\,\{\mathcal{C}, \mathcal{N}, \mathcal{Q}\}\right)
\bigr\},
\label{eq:clean}
\end{equation}
where $\mathcal{C}$ ensures completeness, $\mathcal{N}$ enforces unit-normalization coherence, and $\mathcal{Q}$ restricts to valid quasi-steady-state profiles. The predicate $\mathcal{V}_{\mathrm{GK}}$ operationalizes these quality criteria without binding to any operating point. {Each predicate in $\mathcal{V}_{\mathrm{GK}}$ is verified as follows. 
\textbf{Completeness} ($\mathcal{C}$) ensures that all core variables in Eq.~(\ref{eq:gkfeatures}) are present; 
records missing geometry, gradient, or thermodynamic terms are discarded. 
\textbf{Normalization coherence} ($\mathcal{N}$) checks unit consistency and normalization to machine parameters 
(e.g., $R_0$, $c_s$), reconciling mixed units or gradient definitions. 
\textbf{Quasi–steady-state validity} ($\mathcal{Q}$) requires temporal stability, 
retaining data where relative variations $|\dot{X}/X|<5\%$ over several confinement times. 
Together, these filters ensure physically consistent and reproducible parameter tuples for graph-based analysis.
}

Device-specific parameters are then transformed into the unified feature space \eqref{eq:gkfeatures} via a deterministic formatting operator $F(\cdot)$. {Here, $F(\cdot)$ denotes a deterministic formatting operator that converts heterogeneous, device-specific quantities into the unified feature space defined in Eq.~(\ref{eq:gkfeatures}). It standardizes variable names, applies normalization rules (e.g., to $R_0$, $a$, and $c_s$), and reformats derived quantities such as $R/L_y = -\,(R/y)\,(\partial y/\partial r)$ to ensure consistency across all sources.} For instance, normalized gradient lengths are computed from experimental profile data using
\begin{equation}
\left\{
\begin{aligned}
D_{\mathrm{fmt}} &= F\!\left(D_{\mathrm{clean}}\right), \\
\frac{R}{L_y} &= -\,\frac{R}{y}\,\frac{\partial y}{\partial r},
\end{aligned}
\right.
\label{eq:format}
\end{equation}
where $y \in \{n,\,T_e,\,T_i\}$. All notation and normalization choices conform to the conventions adopted in cross-code GK/transport benchmarks~\cite{bourdelle2015core}.

To facilitate reproducibility and transparency in retrieval and generation, we summarize internal statistics over $\mathcal{X}_{\mathrm{GK}}$ coordinates, which is given by Eq.~(\ref{eq:stats_method})
\begin{equation}
\left\{
\begin{aligned}
\mu &= \tfrac{1}{n}\sum_{i=1}^{n}x_i, \\
\sigma &= \sqrt{\tfrac{1}{n-1}\sum_{i=1}^{n}(x_i-\mu)^2},
\end{aligned}
\right.
\label{eq:stats_method}
\end{equation}
where $x_i$ denotes the $i$-th sample in a given coordinate. These corpus-level statistics support interpretability, normalization, and coverage estimation in the later GraphRAG retrieval process. 

\subsection{LLM with GraphRAG}

\begin{figure*}[!t]       
	\centering
        
        \includegraphics[width=\textwidth]{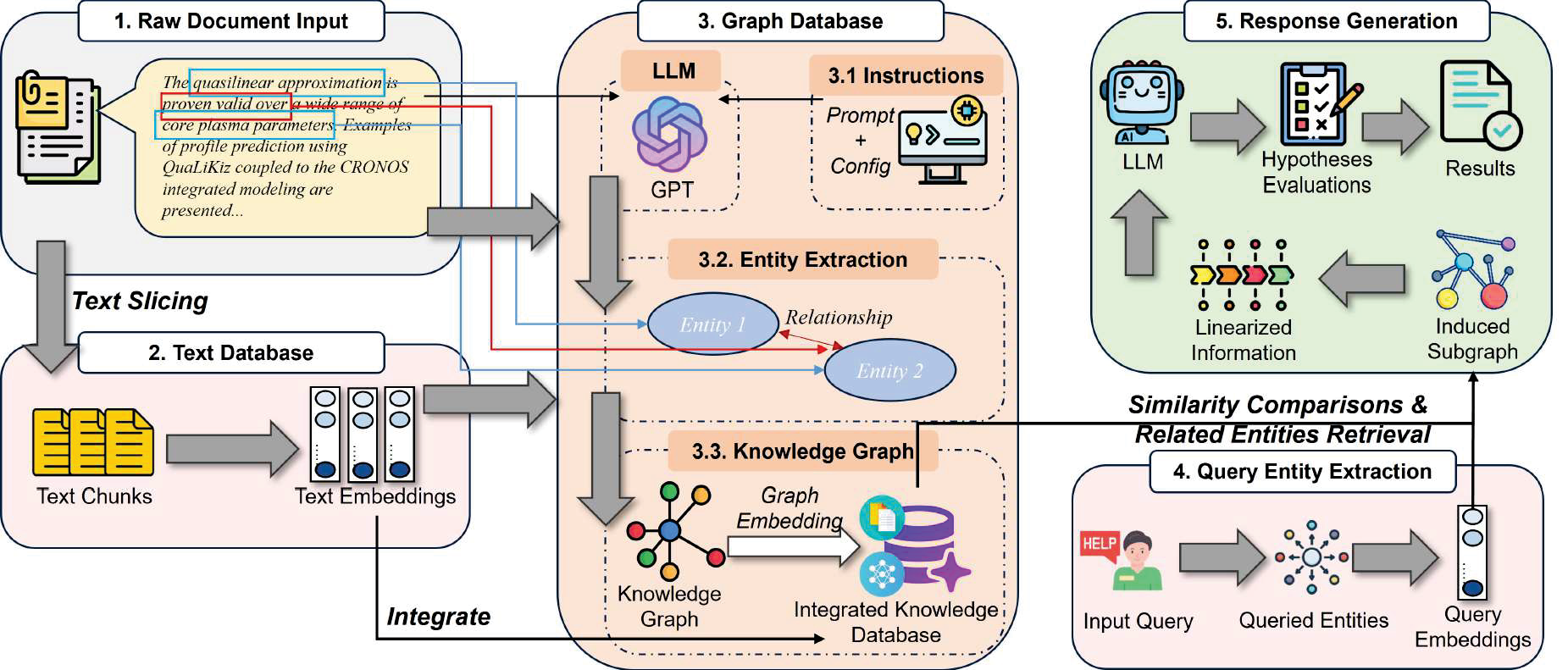}
        
	\caption{LLM-guided parameter range recommendation grounded in structured retrieval.
}   
	\label{FIG:workflow}
\end{figure*}

The second stage of \textbf{Plasma GraphRAG} integrates large language models with structured retrieval over a typed parameter graph tailored for GK simulations. As illustrated in Figure~\ref{FIG:workflow}, {the system processes raw plasma literature into semantic chunks and extracts entities and relations to build a physics-grounded knowledge graph. This procedure is semi-automated: initial entity and relation extraction is performed using a domain-adapted NLP pipeline 
based on named-entity recognition (NER) and dependency parsing, while manual validation and correction are applied to ensure physical consistency and symbol standardization. In principle, the pipeline can operate fully automatically for large-scale ingestion, 
but expert-in-the-loop curation remains essential to maintain accuracy in highly technical contexts.} Given a natural-language query, relevant parameter nodes are retrieved and expanded into a $d$-hop evidence subgraph, which is then linearized and provided to the LLM \cite{wu2024retrieval}. The model generates multiple answer candidates, guided by a reranking objective that promotes evidence coverage and factual consistency. When evidence is insufficient, the system abstains or returns a low-confidence response. This section details each component of the pipeline, including graph indexing, retrieval, linearization, and generation.

\subsubsection{Graph-Based Indexing}

Plasma GraphRAG constructs a typed, text-attributed graph $G=(V,E,T)$ that encodes domain knowledge for gyrokinetic modeling. Nodes $v \in V$ are typed via a map $\tau : V \to \{\mathsf{param}, \mathsf{dev}, \mathsf{src},\ldots\}$, where parameter nodes are defined as
\begin{equation}
P = \left\{\, v \in V \,\middle|\, \tau(v)=\mathsf{param} \right\}.
\label{eq:param_nodes}
\end{equation}
Each parameter node $v \in P$ carries an attribute $x_v = T(v)$ that aggregates definitional text, figure captions, and bibliographic snippets. Prior to attribution, a normalization layer $\mathcal{N}_{\mathrm{sym}}$ (NER and regex-based) standardizes symbols and aliases.

\textbf{{Text Embeddings:}}
Descriptions $x_v$ are embedded using a sentence-level encoder, such as SentenceBERT~\cite{reimers2019sentence}, yielding
\begin{equation}
z_v = \mathrm{LM}(x_v) \in \mathbb{R}^d, \qquad v \in P.
\label{eq:embedding}
\end{equation}
Stacked embeddings form the parameter matrix, i.e.,
\begin{equation}
Z =
\begin{bmatrix}
z_{p_1}^\top \\
z_{p_2}^\top \\
\vdots \\
z_{p_{|P|}}^\top
\end{bmatrix}
\in \mathbb{R}^{|P| \times d}, \quad P = \{p_1, \ldots, p_{|P|}\}.
\label{eq:Zmatrix}
\end{equation}

\textbf{Typed Edges and Weights:}
The overall edge set is the union of relation-specific subsets, i.e.,
\begin{equation}
E = \bigcup_{r \in \mathcal{R}} E^{(r)}, \qquad
\mathcal{R}=\{r_1,r_2,r_3,r_4,\ldots\},
\label{eq:typed_edges}
\end{equation}
with $r_1=\mathsf{co\text{-}mention}$, $r_2=\mathsf{definition\text{-}link}$, 
$r_3=\mathsf{physical\text{-}coupling}$, and $r_4=\mathsf{table\text{-}row}$.
 For parameter–parameter pairs $(p_i, p_j) \in P \times P$, the edge weight is computed as
\begin{equation}
w(p_i, p_j) =
\alpha \, \mathrm{sim}(z_{p_i}, z_{p_j}) +
\beta \, \mathrm{cooc}(p_i, p_j) +
\gamma \, \mathrm{phys}(p_i, p_j),
\label{eq:weight_mix}
\end{equation}
where $\alpha + \beta + \gamma = 1$ and each term captures semantic similarity, co-mention frequency, or documented physical coupling, respectively. The resulting adjacency matrix is given by
\begin{equation}
A_{ij} = w(p_i, p_j), \qquad A \in \mathbb{R}^{|P|\times|P|}.
\label{eq:adjacency}
\end{equation}
Thresholding and top-$k$ sparsification may be applied to derive a binary graph when needed.

\textbf{Normalization and Storage:}
For downstream diffusion and aggregation, a symmetric normalization is applied as
\begin{equation}
\left\{
\begin{aligned}
\widetilde{A} &= D^{-\tfrac{1}{2}} (A + I) D^{-\tfrac{1}{2}}, \\
D &= \mathrm{diag}\!\left((A + I)\mathbf{1}\right).
\end{aligned}
\right.
\label{eq:norm_adj}
\end{equation}
The index stores both the raw and normalized graphs: $(Z, A)$ and $(Z, \widetilde{A})$, which expose semantic structure and domain-specific relations for GraphRAG-based retrieval. This completes the graph-based indexing layer.

\subsubsection{Graph-Guided Retrieval}
\label{subsec:retrieval}

This module selects a query-specific, topology-aware evidence subgraph that captures both semantic relevance and domain-specific relations, serving as the structured context for generation. Given a natural-language query $x_q$, we first embed it using the same sentence encoder employed at indexing time, i.e.,
\begin{equation}
z_q \;=\; \mathrm{LM}(x_q)\ \in\ \mathbb{R}^{d},
\label{eq:query_embed}
\end{equation}

To reflect the compositional nature of scientific queries, we extract salient entities (e.g., parameter names, device features) from $x_q$ and embed each individually, i.e.,
\begin{equation}
z_{e_i} \;=\; \mathrm{LM}(e_i)\ \in\ \mathbb{R}^{d}, 
\qquad e_i \in \mathcal{E}_q,
\label{eq:entity_embed}
\end{equation}
yielding a query representation $\{z_{e_1},\ldots,z_{e_{|\mathcal{E}_q|}}\}$ aligned with graph node embeddings.

Node-query relevance is computed via cosine similarity, i.e.,
\begin{equation}
\mathrm{cos}(u,v) \;=\; \frac{u^{\top}v}{\lVert u\rVert_2\,\lVert v\rVert_2},
\label{eq:cosine}
\end{equation}
with the final score per node $p \in P$ given by the best-aligned entity embedding, i.e.,
\begin{equation}
\mathrm{score}(p) \;=\; \max_{e_i \in \mathcal{E}_q}\ \mathrm{cos}\!\bigl(z_{e_i},\, z_p\bigr).
\label{eq:score_node}
\end{equation}
Top-$k$ parameter seeds are then selected as
\begin{equation}
P_k \;=\; \mathrm{TopK}_{\,p\in P}\ \mathrm{score}(p).
\label{eq:topk}
\end{equation}
Here, the top-$k$ operation selects the $k$ parameter nodes with the highest semantic similarity scores to the query, forming the initial seed set for graph expansion.

To incorporate inductive biases and neighborhood semantics, a composite score is introduced, i.e.,
\begin{equation}
s_{\theta}(p\mid q) \;=\; 
\lambda_1\,\mathrm{cos}\!\bigl(z_q, z_p\bigr)
\;+\; \lambda_2\,\tau(p \parallel q)
\;+\; \lambda_3\,\max_{u\in \mathcal{N}(p)} \mathrm{cos}\!\bigl(z_q, z_u\bigr),
\label{eq:typed_score}
\end{equation}
{where $\lambda_i \ge 0$ and $\sum_i \lambda_i = 1$. 
The weights control the contribution of different relevance components: 
$\lambda_1$ emphasizes direct semantic similarity between the query and parameter node, 
$\lambda_2$ adjusts the influence of the type prior $\tau(p \parallel q)$, 
and $\lambda_3$ accounts for neighborhood similarity by aggregating contextual information from adjacent nodes $\mathcal{N}(p)$. 
The function $\tau(p \parallel q)$ encodes a type prior (e.g., boosting parameters for range-related queries), 
and $\mathcal{N}(p)$ denotes neighbors of node $p$. 
Either the raw semantic score in Eq.~(\ref{eq:score_node}) or the hybrid score $s_{\theta}(\cdot)$ in Eq.~(\ref{eq:typed_score}) may be used to derive $P_k$.}

To recover broader relational context, the selected seed set is expanded via a $d$-hop neighborhood, which is given by
\begin{equation}
\left\{
\begin{aligned}
&P^{\ast} = P_k \,\cup\, \mathcal{N}^{(d)}(P_k), \\
&\mathcal{N}^{(d)}(S) = \cup_{p\in S}\,\mathcal{N}^{(d)}(p),
\end{aligned}
\right.
\label{eq:expansion}
\end{equation}
where $d$ is a tunable hyperparameter.

The final evidence subgraph is the induced subgraph on this expanded node set, i.e.,
\begin{equation}
G^{\ast} \;=\; \mathrm{induce}\!\bigl(P^{\ast},\,E\bigr),
\label{eq:induced}
\end{equation}
which is then passed to the linearization and generation modules. Formally, the retrieval objective can be cast as
\begin{equation}
G^{\ast}\;=\;\arg\max_{\,G' \subseteq \mathcal{R}(G)}\ 
\mathrm{Sim}\!\bigl(x_q,\,G'\bigr),
\label{eq:retrieval_obj}
\end{equation}
where $\mathcal{R}(G)$ is the feasible subgraph region (e.g., those rooted at $P_k$) and $\mathrm{Sim}(\cdot,\cdot)$ aggregates node-level affinities and optional coverage/diversity criteria. The choices of $k$, $d$, and weights $\lambda_i$ are specified in the Experiments section.

\subsubsection{Answer Generation}
\label{subsec:generation}

Conditioned on the retrieved subgraph $G^{\ast}=(P^{\ast},E^{\ast})$, the generator synthesizes responses explicitly grounded in the linearized evidence. The query and subgraph context are concatenated into a single input sequence under a traversal policy $\pi$, ensuring stable ordering and provenance retention, i.e.,
\begin{equation}
F(x_q, G^{\ast}) \;=\; \bigl[\, \mathrm{lin}(G^{\ast};\,\pi)\ ;\ x_q \,\bigr].
\label{eq:compose}
\end{equation}

The generation objective is posed as conditional maximum likelihood over the answer space $\mathcal{A}$, which is given by
\begin{equation}
a^{\ast} \;=\; \arg\max_{a \in \mathcal{A}} \; p_{\phi}\!\bigl(a \mid F(x_q, G^{\ast})\bigr),
\label{eq:gen_objective}
\end{equation}
with autoregressive factorization over tokens. To encourage explicit grounding, decoding is guided by a reranking objective that rewards coverage of retrieved evidence and penalizes hallucinations or verbosity. Among an $N$-hypothesis set $\mathcal{H}$, the final answer $a^{\dagger}$ is selected by maximizing this reranking score \cite{zhao2024retrieval} When retrieval is insufficient, the system abstains from answering or marks the output as low-confidence. Confidence is estimated from the aggregate relevance of the retrieved nodes and the coverage of evidence cited in the response. Post-processing further enforces citation consistency and output formatting to match plasma-physics conventions.

Algorithm~\ref{alg:plasma} summarizes the end-to-end workflow of Plasma GraphRAG. 
Starting from a natural-language query, the system encodes entities, scores parameter nodes, and expands them into a query-specific evidence subgraph. 
This subgraph is linearized and combined with the query to form the input for the generator, which produces multiple candidate answers. 
A reranking mechanism then selects the final output by jointly maximizing likelihood and grounding while penalizing hallucinations and verbosity. 
If retrieval signals are weak, the system abstains or assigns low confidence, thereby maintaining reliability. 
Through this pipeline, Plasma GraphRAG ensures that parameter recommendations are both interpretable and aligned with plasma-physics conventions.

\begin{algorithm}[!t]
\caption{Plasma GraphRAG: GraphRAG-based Parameter Identification}
\label{alg:plasma}
\begin{algorithmic}[1]
\STATE \textbf{Input:} Query $x_q$; graph $G=(V,E,T)$ with parameter nodes $P\subseteq V$ and embeddings $\{z_p\}_{p\in P}$; hyperparameters $k$, $d$, traversal policy $\pi$; rerank weights $(\eta_1,\eta_2,\eta_3)$; thresholds $(\tau,\kappa)$; hypothesis count $N$.
\STATE \textbf{Output:} Answer $a^{\dagger}$ or \textsc{Abstain}

\STATE \textit{// Step 1: Encode query and entities}
\STATE $z_q \leftarrow \mathrm{LM}(x_q)$
\STATE Extract query entities $\mathcal{E}_q=\{e_i\}$ and encode each $z_{e_i}\leftarrow \mathrm{LM}(e_i)$

\STATE \textit{// Step 2: Score candidate parameters}
\FOR{each $p \in P$}
    \STATE $\mathrm{score}(p) \leftarrow \max_{e_i \in \mathcal{E}_q} \mathrm{cos}(z_{e_i}, z_p)$
\ENDFOR

\STATE \textit{// Step 3: Retrieve relevant subgraph}
\STATE $P_k \leftarrow \text{Top-}k$ scored parameters
\STATE $P^{\ast} \leftarrow P_k \cup \mathcal{N}^{(d)}(P_k)$
\STATE $G^{\ast} \leftarrow \text{induced subgraph from } P^{\ast}$

\STATE \textit{// Step 4: Compose input for generation}
\STATE $x_{\mathrm{in}} \leftarrow [\, \mathrm{lin}(G^{\ast};\pi);\ x_q\,]$

\STATE \textit{// Step 5: Generate and rerank answers}
\STATE Generate $N$ hypotheses $\mathcal{H}=\{a^{(1)},\dots,a^{(N)}\}$
\FOR{each $a \in \mathcal{H}$}
    \STATE Compute coverage $\mathrm{cov}(a)$, hallucination penalty $\mathrm{hall}(a)$, length $\mathrm{len}(a)$
    \STATE $\mathcal{J}(a) \leftarrow \log p_{\phi}(a\mid x_{\mathrm{in}}) + \eta_1\,\mathrm{cov}(a) - \eta_2\,\mathrm{hall}(a) - \eta_3\,\mathrm{len}(a)$
\ENDFOR
\STATE $a^{\dagger} \leftarrow \arg\max_{a\in \mathcal{H}} \mathcal{J}(a)$

\STATE \textit{// Step 6: Confidence check}
\STATE $\mathrm{conf} \leftarrow \frac{1}{|P^{\ast}|}\sum_{p\in P^{\ast}} \mathrm{score}(p)$
\IF{$\mathrm{conf} < \tau$ \textbf{or} $\mathrm{cov}(a^{\dagger}) < \kappa$}
    \STATE \textbf{return} \textsc{Abstain}
\ELSE
    \STATE \textbf{return} $a^{\dagger}$
\ENDIF
\end{algorithmic}
\end{algorithm}

\begin{figure*}[!t]       
	\centering
        \includegraphics[width=0.9\textwidth]{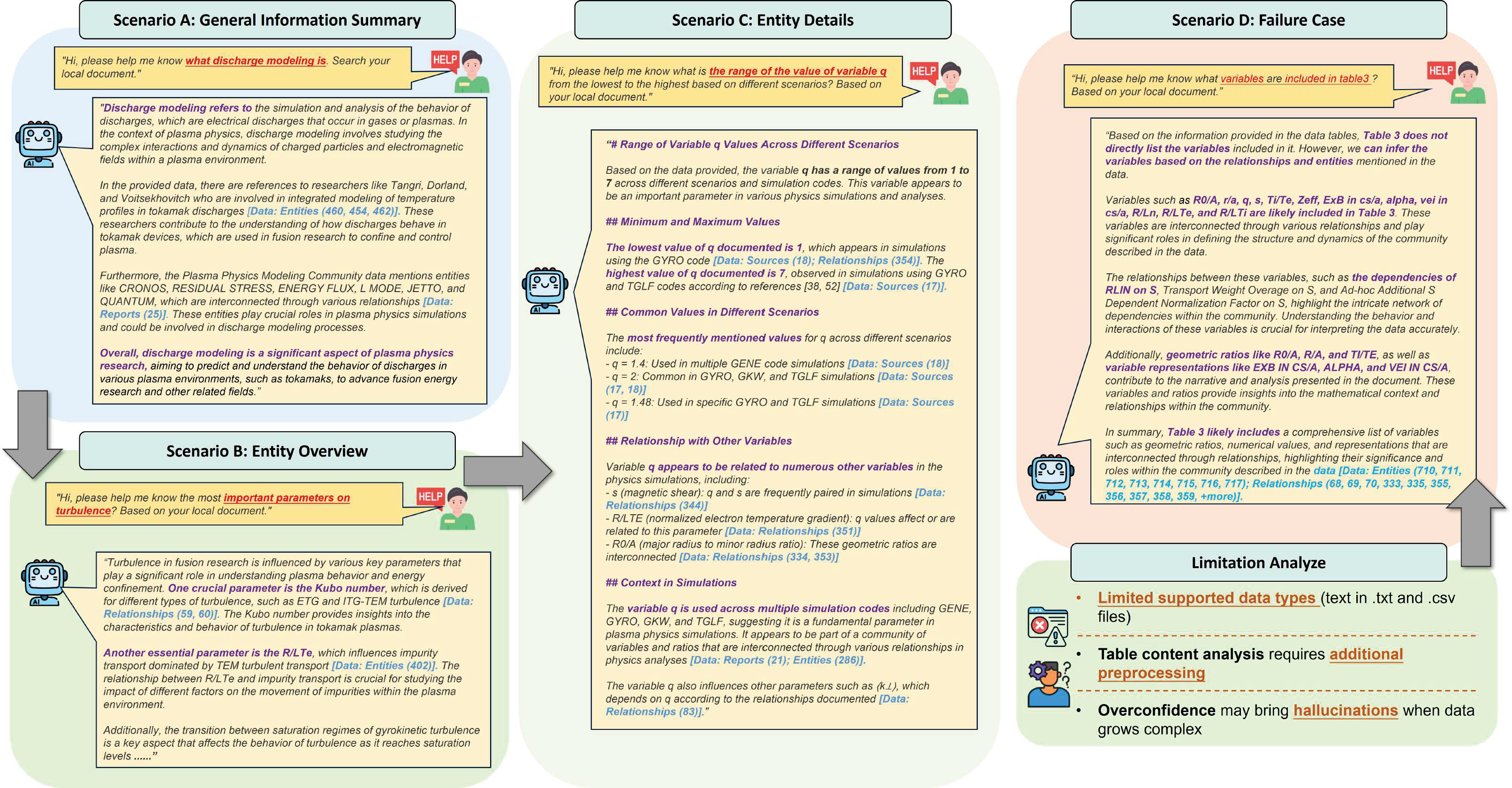}
	\caption{Visualization of sample user interactions with the Plasma GraphRAG and sample response generations. {The answer is generated using GPT-4o as the generator.}
}   
	\label{FIG:0}
\end{figure*}

\section{Numerical Results}

\subsection{Parameter Setting}

{To assess the effectiveness of \textbf{Plasma GraphRAG} in physics-grounded parameter identification, we constructed a controlled question-answering benchmark targeting key aspects of gyrokinetic modeling. The evaluation set comprises 10 representative questions drawn from canonical literature and simulation benchmarks, spanning four categories: (i) equilibrium and geometry descriptors, (ii) thermodynamic ratios and species composition, (iii) transport-driving gradients, and (iv) stability and collisionality proxies. All questions are posed in natural language and require extraction or inference of parameter ranges grounded in the source corpus.}

We compare three models under identical retrieval configurations: (1) GraphRAG with \texttt{GPT-3.5-turbo}, (2) GraphRAG with \texttt{LLaMA-3.1-8B}, and (3) a vanilla RAG baseline. The corpus consists of peer-reviewed gyrokinetic studies normalized according to established conventions and encoded into a text-attributed, typed parameter graph for structured retrieval. {We set the retrieval chunk size to 1200 tokens with 100 tokens overlapping to ensure the construction of the knowledge graph captures sufficient entities and relationships while retaining enough context information for interpreting them properly.} Responses are evaluated using five metrics: \emph{Diversity}, \emph{Comprehensiveness}, \emph{Hallucination}, \emph{Directness}, and \emph{Empowerment}. {These metrics respectively assess coverage breadth, factual grounding, linguistic clarity, and practical usefulness, with hallucination defined as any statement inconsistent with the retrieved subgraph}~\cite{yu2024evaluation}. {All evaluations use a temperature of 0.0 for deterministic outputs. We validate reproducibility by running each evaluation three times.}

\subsection{Simulation Results}

As illustrated in Figure~\ref{FIG:0}, we present a case study highlighting how Plasma GraphRAG facilitates grounded interactions for gyrokinetic parameter exploration. In Scenario A, the user seeks general knowledge about discharge modeling; the system retrieves related entities, documents, and researchers, offering a comprehensive and citation-backed summary. Scenario B and C delve into parameter-specific queries—identifying key variables for turbulence and quantifying the range of a specific parameter across simulations. In both cases, the agent retrieves and linearizes a relevant evidence subgraph from the knowledge graph, enabling responses that are both precise and interpretable. This showcases the agent’s ability to support nuanced scientific inquiry through structured retrieval, minimizing hallucination while promoting traceability and domain consistency.

\begin{figure}[!t]       
	\centering
        
        \includegraphics[width=0.49\textwidth]{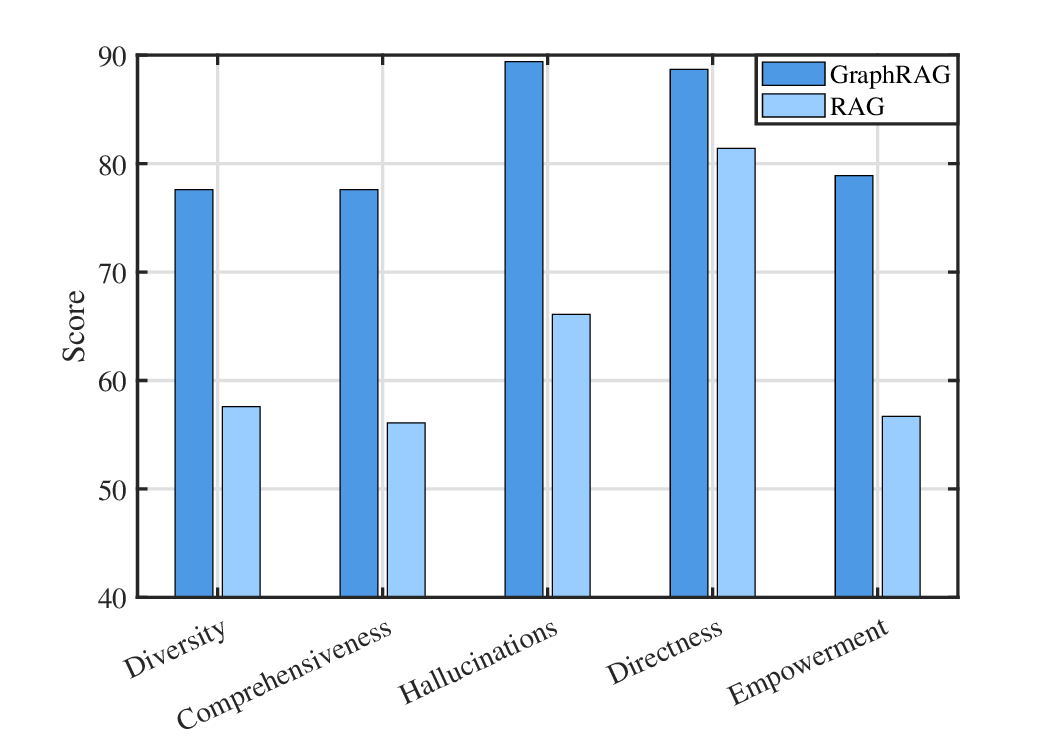}
        
	\caption{Experiment results for comparing performance between GraphRAG and Vanilla RAG with GPT-4o. 
}   
	\label{FIG:1}
\end{figure}

Figure~\ref{FIG:1} compares GraphRAG with GPT-4o against the vanilla RAG baseline across five evaluation metrics. GraphRAG achieves consistently higher scores {in all categories}, demonstrating its ability to capture a broader range of plasma parameters and deliver responses that are both informative and actionable for simulations.

{Diversity measures the variety of unique parameters correctly mentioned in each answer, while comprehensiveness quantifies the proportion of relevant parameters covered relative to ground-truth literature references. Hallucination captures the percentage of statements unsupported or contradicted by the retrieved subgraph, serving as an indicator of factual reliability. Directness reflects linguistic clarity and conciseness, computed as the inverse of average response length normalized by informativeness, and empowerment evaluates how actionable or simulation-ready the suggested parameter ranges are, as judged by expert annotators.
All scores are normalized to a 0–100 scale and averaged over ten benchmark queries}.

It also shows a {$35.25\%$ reduction} in hallucinations, confirming that graph-structured retrieval helps ground responses more faithfully in the literature. Overall, the results highlight that GraphRAG provides a more balanced and reliable framework, better suited for accuracy, reproducibility, and interpretability in plasma parameter determination.

\begin{figure}[!t]       
	\centering
        
        \includegraphics[width=0.49\textwidth]{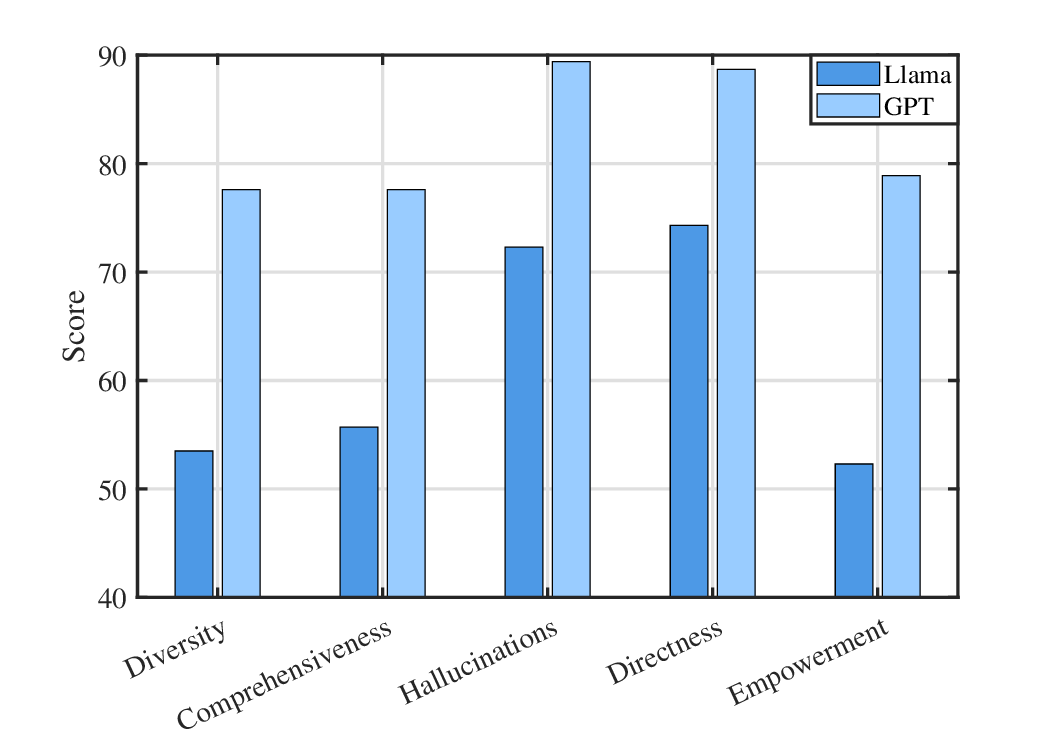}
        
	\caption{Experiment results for comparing performance between GraphRAG with Llama3.1-8b and GraphRAG with GPT-4o.
}   
	\label{FIG:2}
\end{figure}

Figure~\ref{FIG:2} compares the performance of GraphRAG when paired with {Llama3.1-8B} and GPT-4o across five evaluation metrics, where higher scores indicate better performance. The results show that GPT outperforms Llama across all metrics, demonstrating its superior ability to generate broad, accurate, and well-grounded parameter recommendations. This improvement highlights GPT’s stronger reasoning capacity and richer contextual understanding, which enable it to capture complex relationships among gyrokinetic parameters and produce more informative and actionable responses. In contrast, Llama delivers comparatively narrower and less detailed outputs, reflecting its limited contextual modeling capability. Overall, GPT provides the most balanced and high-quality performance across all evaluation dimensions.

\begin{figure}[!t]       
	\centering
        \includegraphics[width=0.49\textwidth]{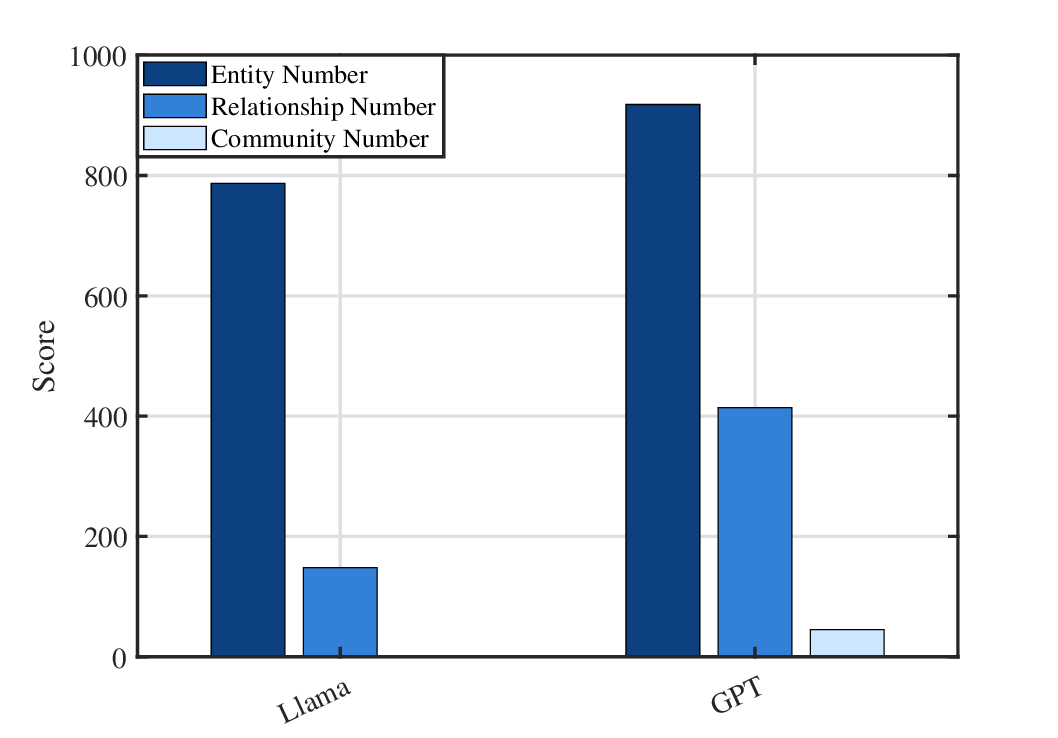}
	\caption{Components in the Knowledge Graph constructed with Llama3.1-8b and GPT-3.5-turbo. 
}   
	\label{FIG:3}
\end{figure}

Figure~\ref{FIG:3} compares the structural components of the knowledge graphs constructed with Llama and GPT-3.5-turbo, focusing on the number of entities, relationships, and detected communities. While both models extract a large set of entities from the plasma literature, GPT identifies more entities (918 vs. 787) and, more importantly, captures a much higher number of relationships (414 vs. 148). This richer connectivity translates into a graph with stronger inter-parameter links, which provides the retrieval model with better context for answering parameter-related queries. {Moreover, GPT detects 45 distinct communities within the knowledge graph, whereas Llama yields only a single loosely connected cluster. Each community corresponds to a cohesive subgraph in which entities co-occur frequently across the literature and share strong semantic or physical associations. In practice, these clusters align closely with meaningful physics concepts. For example, one community centers on magnetic geometry descriptors, another on turbulence-driving gradients, and others on collisionality or shearing rates. This structure indicates that GPT not only captures a denser web of parameter relationships but also organizes them into interpretable, physics-consistent domains. Such emergent clustering enhances both the transparency and the interpretability of downstream GraphRAG reasoning, enabling the agent to retrieve evidence that mirrors the way plasma physicists naturally group related quantities.}

\begin{figure}[!t]       
	\centering
        \includegraphics[width=0.49\textwidth]{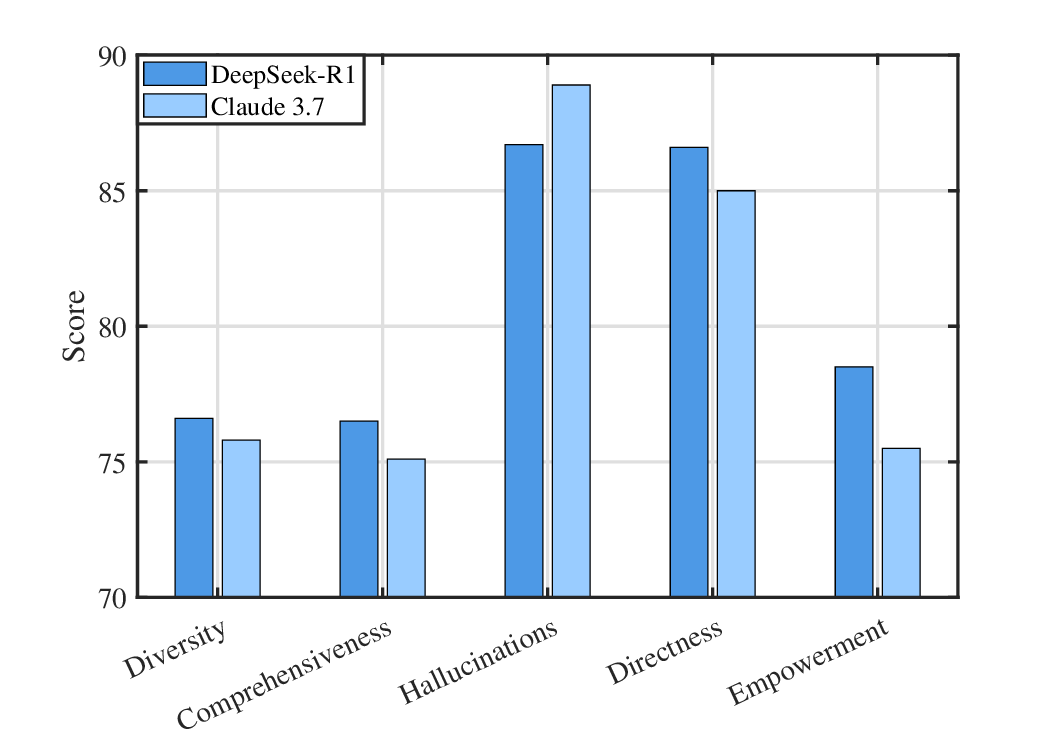}
	\caption{Experiment results for comparing performance between GraphRAG with DeepSeek-R1 and GraphRAG with Claude 3.7 Sonnet.
}   
	\label{FIG:4}
\end{figure}

Figure~\ref{FIG:4} compares the performance of GraphRAG when combined with {DeepSeek-R1} and {Claude 3.7 Sonnet} across the five evaluation metrics, where higher scores indicate better performance. Overall, DeepSeek-R1 achieves consistently higher or comparable scores in all metrics, showing clear advantages in \emph{comprehensiveness}, \emph{hallucination control}, and \emph{empowerment}. This suggests that DeepSeek-R1 produces broader, more reliable, and practically useful parameter recommendations. Claude 3.7, on the other hand, performs slightly better in \emph{directness} and maintains competitive diversity, indicating that its responses are concise and well-structured but somewhat less extensive in contextual coverage. Taken together, the results demonstrate that DeepSeek-R1 offers more balanced and overall stronger performance, while Claude 3.7 prioritizes brevity and clarity in its outputs.

\begin{figure}[!t]       
	\centering
        \includegraphics[width=0.49\textwidth]{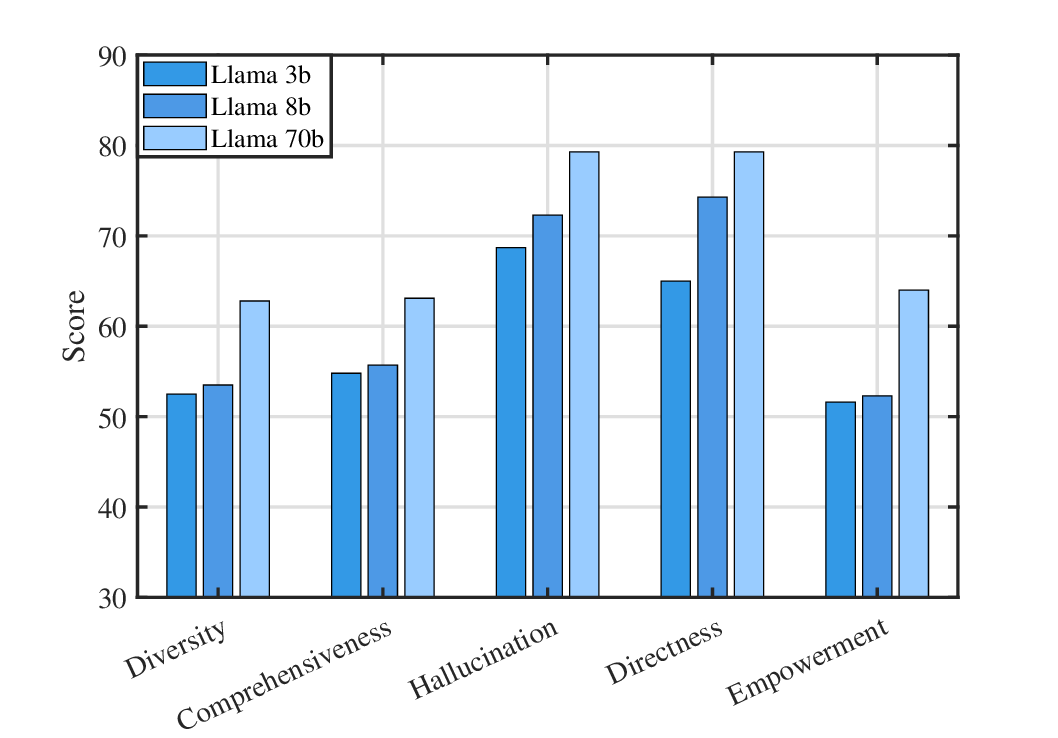}
	\caption{Scores of Llama models (3B, 8B, 70B) across five evaluation metrics, showing clear gains with larger model sizes.
}   
	\label{FIG:5}
\end{figure}

Figure~\ref{FIG:5} presents the evaluation of three Llama models with increasing parameter sizes, 3B, 8B, and 70B, across the five performance metrics. The results show a clear positive correlation between model scale and overall performance. Llama-70B achieves the highest scores in almost all metrics, particularly in directness and hallucination control, where it approaches 80 points, indicating that larger models not only provide clearer and more precise answers but also remain more faithful to the retrieved evidence. Improvements are also evident in diversity and comprehensiveness, suggesting that the expanded capacity of the 70B model enables it to capture a wider range of plasma descriptors and generate richer, more context-aware parameter recommendations. By contrast, the smaller Llama-3B and Llama-8B models perform considerably lower, especially in empowerment, reflecting their limited ability to produce guidance that is practically useful for parameter setting.

\begin{figure}[!t]       
	\centering
        \includegraphics[width=0.49\textwidth]{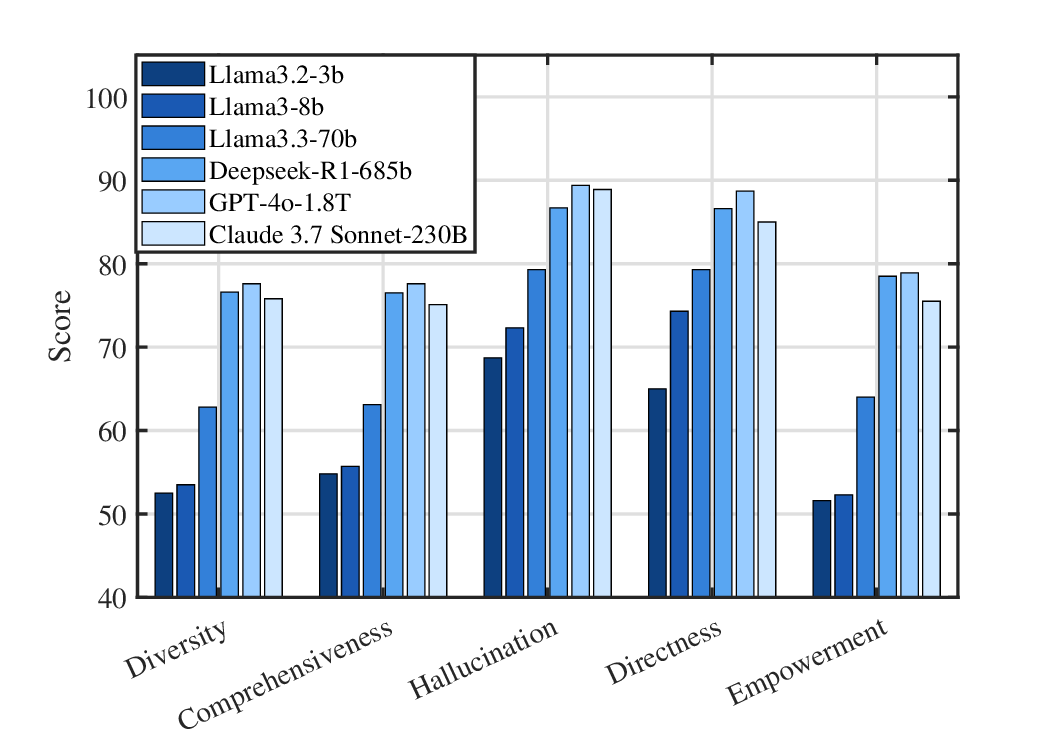}
	\caption{Comparison of multiple LLM families, where ultra-large models (GPT-4o, Claude) outperform smaller Llamas and DeepSeek-R1 shows strength in diversity.
}   
	\label{FIG:6}
\end{figure}
Figure~\ref{FIG:6} extends the evaluation to a broader set of LLM families, including the Llama series (3B, 8B, 70B), DeepSeek-R1 (685B), GPT-4o (1.8T), and Claude 3.7 Sonnet (230B). The results reveal a clear scaling trend: while smaller Llama models achieve only modest performance across all metrics, larger-scale systems demonstrate dramatic improvements, particularly in comprehensiveness, directness, and empowerment. Among the ultra-large models, GPT-4o and Claude 3.7 Sonnet dominate the evaluation, reaching near-perfect scores in comprehensiveness and maintaining strong performance in diversity and empowerment, which underscores their superior reasoning and grounding capacity. DeepSeek-R1 also performs competitively, especially in diversity and hallucination control, reflecting its architectural emphasis on deep reasoning. In contrast, even the largest Llama-70B lags behind these frontier models, showing that while scaling within a family improves results, architecture and training quality remain decisive factors. Overall, the figure highlights a hierarchy in capability: smaller Llamas are lightweight but limited, while ultra-large models like GPT-4o and Claude set the benchmark for high-quality, well-grounded responses, albeit at much higher computational cost.

\section{Conclusion}
In this work, we have introduced Plasma GraphRAG, a framework that integrates GraphRAG with large language models to automate the identification of parameter ranges in gyrokinetic simulations. Unlike traditional manual reviews, Plasma GraphRAG constructs a physics-informed knowledge graph and applies structured retrieval to explicitly capture parameter relationships. This design enables accurate, comprehensive, and reproducible recommendations while reducing hallucinations. Experimental results show that GraphRAG consistently outperforms vanilla RAG across key metrics such as diversity, grounding, and interpretability. {Nevertheless, the current evaluation is limited by the relatively small benchmark dataset and the use of heuristic metrics for assessing output quality. Future work will expand the benchmark to cover a broader range of plasma regimes and simulation codes, incorporate quantitative validation against experimental data, and explore reinforcement learning–based optimization for adaptive retrieval and evidence weighting. Overall, Plasma GraphRAG accelerates surrogate model development and provides a scalable foundation for reliable, interpretable parameter selection in plasma physics and other scientific domains.}

\bibliographystyle{IEEEtran}
\bibliography{references}

\end{document}